\documentclass[12pt]{article}

\usepackage{epsfig}

\usepackage{bm}

\newcommand{\eq}{\begin{equation}}
\newcommand{\eqx}{\end{equation}}
\newcommand{\eqn}{\begin{eqnarray}}
\newcommand{\eqnx}{\end{eqnarray}}

\begin{document}

\date{}

\title{On the SO(9) structure \\of supersymmetric Yang-Mills quantum mechanics}

\author{
J. Wosiek \\
M.Smoluchowski Institute of Physics, Jagellonian University\\
Reymonta 4, 30-059 Krak\'{o}w, Poland}

\maketitle

\begin{abstract}
In ten space-time dimensions the number of Majorana-Weyl fermions is not conserved, not only during the time
evolution, but also by rotations. As a consequence the empty Fock state is not rotationally symmetric. We
construct explicitly the simplest singlet state which provides the starting point for building up invariant
SO(9) subspaces. The state has non-zero fermion number and is a complicated combination of the 1360
elementary, gauge invariant, gluinoless Fock  states with twelve fermions. Fermionic structure of higher
irreps of SO(9) is also briefly outlined.
\end{abstract}
\newline

\vspace*{1cm}
\noindent TPJU-4/2005 \newline
March 2005 \newline
\newpage
\section{Introduction}

Supersymmetric Yang-Mills Quantum Mechanics (SYMQM) emerge from the dimensional reduction
of corresponding field theories to a single point in the D-1 dimensional space \cite{CH,BR}. Resulting
systems are characterized by two parameters: D - the dimensionality of the space-time of the unreduced theory, and
N - the number of colors specifying a gauge group.
Dependence on the gauge coupling follows from
the simple rescaling of a finite number of degrees of freedom. The whole family (for various D and N) reveals
a broad range of very interesting phenomena and has many applications in seemingly distant areas of theoretical
physics \cite{WI}-\cite{ST}. Perhaps the most known example is the conjectured relevance of the D=10 system, at large N,
to the M-theory \cite{BFSS,BS,WAT}.

  A series of results has been recently obtained for the D=2 and D=4, SU(2), models
with the aid of the cut Fock space approach \cite{JW1,CW1,CW2}.
In this paper we address the D=10, N=2
system \cite{HS} and construct explicitly the SO(9) singlet state which replaces the empty Fock state sufficient
in lower dimensions. This state is highly non-trivial
due to the non-conservation of the Majorana-Weyl fermion number in ten dimensions.

The hamiltonian reads \cite{HS}
\begin{eqnarray}
H &=& H_K + H_P + H_F, \nonumber \\
H_K &=& {1\over 2} p_a^ip_a^i , \nonumber \\
H_P &=& {g^2\over 4}\epsilon_{abc}
\epsilon_{ade}x_b^i x_c^j x_d^i x_e^j, \label{eq:Hamiltonian} \\
\label{HD4} H_F &=& {i g \over 2}
\epsilon_{abc}\psi_a^{\dagger}\Gamma^k\psi_b x_c^k. \nonumber
\end{eqnarray}
There are 27 bosonic coordinates $x^i_a$, and their momenta $p^i_a$, $i=1,\dots ,9$, $a=1,2,3$.
Fermionic degrees of freedom compose a Majorana-Weyl spinor in the addjoint
representation of the SU(2) gauge group, $\psi_a^{\alpha}$, $\alpha=1,...,16$.
$\Gamma^k$ are 16x16 subblocks of the big (32x32) Dirac $\alpha^k$ matrices in chiral representation. In all
explicit calculations we use the representation of Ref.\ \cite{POL}.

The system has the internal Spin(9) rotational symmetry
generated by the gauge invariant angular momentum
\begin{equation}
J^{kl}=
\left(x^{[k}_a p^{l]}_a+{1\over 2}\psi^{\dagger}_{a}\Sigma^{kl}\psi_{a}\right),
\label{JD10}
\end{equation}
with
\begin{equation}
\Sigma^{kl}=-{i\over 4}[\Gamma^{k},\Gamma^{l}].
\end{equation}
After the dimensional reduction, the local gauge invariance
amounts to the global invariance under the SU(2) rotations generated by
the color angular momentum
\begin{equation}
G_a=\epsilon_{abc}\left(x_b^k p_c^k-{i\over 2}\psi^{\dagger}_b\psi_c \right).
\label{GG}
\end{equation}
Furthermore, the hamiltonian, Eq.(\ref{eq:Hamiltonian}), is invariant under, ${\cal N}=1$, ten dimensional supersymmetry with 16 Majorana-Weyl
generators
\begin{equation}
Q_{\alpha}=(\Gamma^k\psi_a)_{\alpha}p^k_a + i g
\epsilon_{abc}(\Sigma^{jk}\psi_a)_{\alpha}x^j_b x^k_c.
\label{QD4}
\end{equation}
Supersymmetry requires imposing both Weyl and Majorana conditions on the 32 dimensional spinor.
To identify explicitly fermionic degrees of freedom, we construct big (32x32) Dirac $\alpha$ matrices
following Ref.\ \cite{POL}. In chiral representation they are block diagonal, hence we restrict
ourselves to one chirality \footnote{Compared to Ref.\ \cite{POL} simple similarity transformation
is required to bring big $\alpha$ matrices to this form.}. It is well known that both Majorana and Weyl
conditions can be simultaneously imposed only in $D=2 (mod\; 8)$ space-time dimensions. Consequently the Majorana
matrix turns out to be block diagonal as well, and we can impose Majorana condition in one chirality subblock.
Finally, a solution of the Majorana condition, in chiral representation, has a simple form
\begin{equation}
\psi_a^T=(f^1,f^2,f^3,f^4,f^5,f^6,f^7,f^8,{f^8}^{\dagger},-{f^7}^{\dagger},{f^6}^{\dagger},-{f^5}^{\dagger},
-{f^4}^{\dagger},{f^3}^{\dagger},-{f^2}^{\dagger},{f^1}^{\dagger})_a \label{eqweyl}
\end{equation}
with $f^{\dagger}$ and $f$ being the standard, anticommuting, fermionic creation and annihilation operators.
Therefore, the ten dimensional supersymmetric Yang-Mills quantum mechanics, with SU(2) gauge group has
24 fermionic degrees of freedom.
\section{The cut Fock space approach}
There exists surprisingly powerful method to compute the complete spectrum and eigenstates
of polynomial hamiltonians with a "reasonably" large number of degrees of freedom \cite{JW1}.
It was applied successfully to the D=2 and D=4 SYMQM with 6 and 15 degrees of freedom respectively \cite{CW1,CW2}.
In the latter case many new properties of this system were uncovered, including identification of
dynamical supermultiplets, computation of their energies, wave functions, etc.  Ten dimensional system
can also be attacked with this approach. Presumably the high accuracy of Ref. \cite{CW2} could not be matched
at the moment, but the recursive technique developed there offers a real possibility for some quantitative
results. However, the ten dimensional system is more complex, also on the more fundamental level. Namely, the fermion
number is not conserved by the hamiltonian, and also by the SO(9) rotations, Eq.(\ref{JD10}). In this work we
address this difficulty in some detail and propose one possible solution.

In order to better illustrate the problem, we briefly sketch the method of Refs.\ \cite{JW1,CW1,CW2}.
Since the hamiltonian, Eq.(\ref{eq:Hamiltonian}) is a simple function of creation and annihilation operators,
it is convenient use the eigenbasis of the number operators associated with all individual degrees of freedom.
Beginning with the empty (fermionic and bosonic) state, $|0>=|0_F, 0_B>$,
we construct the physical, i.e. gauge invariant, basis of the Hilbert
space by acting on $|0>$ with gauge invariant polynomials of all creation operators.
The basis is artificially cut by limiting
the total number of all bosonic quanta. Then we calculate analytically matrix representation of the hamiltonian
and obtain numerically the spectrum. The procedure is then repeated for higher cutoffs until the
results converge. Originally the analytical part of the calculation was done in Mathematica by defining the
Mathematica representation of the Fock states and all relevant operators. In Ref.\ \cite{CW2} this was replaced
by the fast, recursive calculation of all matrix elements. Where possible conservation laws were used.
For example, the fermionic number is conserved in D=2 and D=4 models and
 the whole procedure was
carried out independently in each fermionic sector. In the four dimensional model we reduced
the problem even further by using (composite) creators with
fixed fermion number and angular momentum. This allowed to obtain the energies
 of the first 10-20 states, in every $J$ $( 0 \le J \le\;\; \sim 16 )$ sector, with the four digit precision.

In ten dimensions fermion number is not conserved \cite{HS}, and things require more care.
Even before that, it turns out that the empty Fock state - the state which is the root of the whole
construction - has to be considerably modified.

\section{A puzzle}

    In dimensionally reduced theories the total number of fermionic quanta is finite. Since all quanta occupy
the same point in space, one can only have as many fermionic quanta as there are different degrees of freedom.
Hence in D=2, there are maximally three, in D=4 -- six, and in D=10 -- 24 fermions. Correspondingly, there are four,
seven and 25 fermionic sectors in the Hilbert spaces of these systems.
In ten dimensions, fermion number is not
conserved, nevertheless we shall use the concept of fermionic sectors just for the purpose
of initial
classification. In each fermionic sector we then construct a complete basis of gauge invariant states allowing
up to $B_{max}$ bosonic quanta. To this end one needs the empty Fock state $|0>=|0_F,0_B>$
mentioned earlier.
In D=2 and 4 this is a simplest possible state, in particular in D=4 it is rotationally symmetric, i.e.
it is annihilated by the angular momentum. This is not the case for D=10. We obtain
\eq
J^2|0>=78|0>.
\eqx
The empty Fock state is an eigenstate of $J^2$, but it is not a singlet! It belongs to some higher
representation of SO(9). Since SO(9) has rank 4, one needs eigenvalues of other three Casimir operators
to identify uniquely this representation. The precise answer is not relevant here. The lowest candidate
has dimensionality 132132 and, in the Dynkin notation \cite{SL}, is labelled by (1120).

    Since the hamiltonian (\ref{eq:Hamiltonian}) respects the SO(9) symmetry, we would like to
construct separate bases in each channel of SO(9) angular momentum.
 To this end however, we need to begin with the
SO(9) singlet state. Evidently the empty state cannot be used for this purpose.
Where is the simplest singlet state? This question will be answered in the next section.

\section{An answer}
Table \ref{bastab} gives sizes of the gauge invariant bases in all 25 fermionic sectors with none and one bosonic quantum.
They were calculated with a fermionic variant of the TWS method introduced in Ref.\ \cite{JW3}\footnote{In short,
we generated all states built from elementary, gauge invariant "bricks":
${f^i_a}^{\dagger} {f^j_a}^{\dagger}$.
Obvious linear dependences  were excluded from the beginning, remaining ones were identified by Gram determinants.
Many states mix, but in a regular pattern. TWS refers to a particular classification scheme, where
the mixings are very transparent and easy to control.}.
They satisfy supersymmetric relations:
 $\#$ bosonic states = $\#$ fermionic states, for each cutoff. Such relations were also found
in lower dimensions within our regularization. It is clear that
the brute force diagonalization of the hamiltonian is out of question, especially for higher $B$ \footnote{For D=4
we needed $B\sim 10-20$ to reach convergent results.}. As discussed earlier, one has to split these bases
into sectors with fixed SO(9) angular momentum.
Hence, again, one needs to construct the simplest singlet state to begin with.

  \begin{table}
  \begin{center}
   \begin{tabular}{cccccccc} \hline\hline
  $F$ &     $ 0 $       &      $ 2 $     &     $ 4 $       &         $ 6 $       &    $ 8 $    & $ 10 $       & $ 12 $        \\  \hline
  $B$ &  $ N_{s}$\ \ & $ N_{s} $ \     &  $ N_{s} $ \ \ & $ N_{s}\ $  \ \ & $ N_{s}\ $  \ \ & $ N_{s}\ $  \ \ & $ N_{s}\ $  \ \     \\
   \hline
  0 &      1\      \ &   28 \      \  &  406  \       \  &     4 060\        \  &  17 605\   \ & 41 392 \ & 56 056     \\
  1 &      -\      \ &  324  \     \  & 9 072 \       \  &  81 648\        \  &     374 544  \    \ &  908 460         \ &   1 205 568            \\
  2 &     45\      \ &  3 816 \    \  & 89 838 \       \  &         \        \  &      \    \ &            \ &             \\
  3 &     84\      \ &  23 652 \   \  &       \       \  &         \        \  &      \    \ &            \ &             \\
  4 &   1 035\      \ &      \     \  &       \       \  &         \        \  &      \    \ &            \ &         \\
  5 &   2 772\      \ &      \      \  &      \       \  &         \        \  &      \    \ &            \ &         \\
  6 &  16 215\      \ &      \      \  &      \       \  &         \        \  &      \    \ &            \ &         \\
   \hline\hline
   \end{tabular}
   \begin{tabular}{rcccccc} 
  $F$ &     $ 1 $       &      $ 3 $     &     $ 5 $       &         $ 7 $       &    $ 9 $    & $ 11 $            \\  \hline
  $B$ &  $ N_{s}$\ \ & $ N_{s} $ \     &  $ N_{s} $ \ \ & $ N_{s}\ $  \ \ & $ N_{s}\ $  \ \ & $ N_{s}\ $   \ \     \\
   \hline
  0 &        -\      \ &   120  \      \  &  1 512\       \  &   8 856\        \  & 29 512 \   \ & 51 520      \\
  1 &       72\      \ &  2 016 \     \  & 29 232 \       \  &  192 528 \        \  &   626 040   \    \ &     1 126 944                  \\
  2 &      288\      \ & 21 024 \    \  &         \       \  &         \        \  &      \    \ &                     \\
  3 &    3 240\      \ &        \   \  &       \       \  &         \        \  &      \    \ &                     \\
  4 &   12 960\      \ &      \     \  &       \       \  &         \        \  &      \    \ &                 \\
  \hline\hline
   \end{tabular}
  \end{center}
\caption{Sizes of bases generated in each (B,F) sector for the
$D=10$ system. $N_s$ is the number of basis vectors.}
\label{bastab}
\end{table}

Since our cutoff
\eq
\Sigma_{b,i} (a^i_b)^{\dagger} a_b^i \le B_{max},
\eqx
is invariant under SO(9) rotations, we can restrict the search to the simplest, $B=0$, sector.
Even then
the problem would require calculation of the huge matrix representation of the complicated $J^2$ operator,
which is practically impossible.

Instead, we have analyzed the action of the 36 components of the angular momentum on all basis states with
$B=0$. Using the explicit representation of Dirac matrices, discussed in Sect.1, we have found that the
four generators from the Cartan subalgebra of SO(9) are particularly simple for $B=0$.
\eqn
J^{23}&=&\frac{1}{2}(N_1-N_2+N_3-N_4+N_5-N_6+N_7-N_8),\nonumber \\
J^{45}&=&\frac{1}{2}(N_1+N_2-N_3-N_4+N_5+N_6-N_7-N_8),\label{car}\\
J^{67}&=&\frac{1}{2}(N_1+N_2+N_3+N_4-N_5-N_6-N_7-N_8), \nonumber \\
J^{89}&=&\frac{1}{2}(N_1+N_2+N_3+N_4+N_5+N_6+N_7+N_8-12),\nonumber
\eqnx
with $N_m=\Sigma_a (f_a^m)^{\dagger}f_a^m$ being the gauge invariant number operator of the m-th Majorana-Weyl
fermion, $m=1\dots 8$. Clearly, above Cartan generators are diagonal in the occupation number representation.
Their eigenvalues can be just read off from the indices of our basis. This substantially simplifies the search
for a spherically symmetric state. First, it is obvious from Eq.(\ref{car}) that a singlet can be only in the sector
with $F=12$. This is the most complex sector of the whole theory with 56056 basis states,
  cf. Table 1. Second,
using again Eqs(\ref{car}) one can readily identify the subset ${\cal B}_{12}(0,0,0,0)$ of the $F=12$ states
with all four magnetic quantum numbers equal to zero. It contains "only" 1360 states. Singlet states, if any,
must be linear combinations of these states. This problem is manageable with our Mathematica representation
of quantum mechanics. We have calculated matrix representation of $J^2$ in the ${\cal B}_{12}(0,0,0,0)$
sub-basis and found the spectrum. It turns out that there exists only one eigenstate with zero eigenvalue.
This is the desired singlet. All others 1359 eigenvalues belong to the known spectrum of the first
Casimir operator of the SO(9), which provides additional check on the whole procedure.
As another test,
we have reconstructed, from the numerical eigenvector, the singlet state in the Fock space,
and checked that, indeed, it is annihilated by all 36 components of the angular momentum, Eq.(\ref{JD10}).

     This is the main result of present paper. The simplest SO(9) invariant state is the linear combination of
1360 basis states from the sector with 12 Majorana-Weyl fermions and no bosonic quanta. Expansion coefficients
are known numerically. This state should be used as the root when creating SO(9) invariant subspaces.
It replaces the empty state of the lower dimensional models. This is one more consequence of the
peculiar behavior of the Majorana-Weyl fermions under SO(9) rotations. Had we not imposed Majorana condition,
the Weyl spinor in Eq.(\ref{eqweyl}) would have consisted of 16 annihilation operators.
 Then the empty state would be symmetric again, cf. Eq.(\ref{JD10}).

      Another way this peculiarity shows up is the following. A fundamental fermionic
representation, (0001),
of SO(9) is 16-dimensional. On the other hand there are only 8 independent
Majorana-Weyl creation operators
\footnote{Clearly
the eight component object built only from creation operators does not form irreducible representation
of SO(9) \cite{SMK}.}.
Therefore we need both creation and annihilation operators to form a covariant SO(9) spinor
as in Eq.(\ref{eqweyl}). Such a spinor, when acting on a singlet state, would create 16 states from
(0001). With the singlet in the $F=12$ sector, everything is consistent: 8 of above states are in
the $F=11$ sector and another 8 have $F=13$. This would have not worked if a singlet was empty.

  \begin{table}
  \begin{center}
   \begin{tabular}{cccccccccccccccc} \hline\hline
  $F$ & $0$ & $2$ & $4$ & $6$ & $8$ & $10$ & $12$  & $14$ & $16$ & $18$ & $20$ & $22$ & $24$ &   irrep   &  dim   \\
  \hline\hline
      &     &     &     &     &     &      & $\times$   &      &      &      &      &      &      & (0,0,0,0) &  1      \\
      &     &     &     &     &     &   $\times$  &  $\times$   &   $\times$  &      &      &      &      &      & (2,0,0,0) &  44     \\
      &     &     &     &     &  $\times$  &   $\times$  &   $\times$   &   $\times$  &  $\times$   &      &      &      &      & (0,0,1,0) &  84     \\
      &     &     &     &  $\times$  &  $\times$  &   $\times$  &   $\times$   &   $\times$  &  $\times$   &  $\times$   &      &      &      & (0,1,1,0) &  1650      \\
      &  $\times$  &  $\times$  &  $\times$  &  $\times$  &  $\times$  &   $\times$  &   $\times$   &   $\times$  &  $\times$   &  $\times$   &  $\times$   &   $\times$  &   $\times$  & (1,1,2,0) &  132132      \\

         \hline\hline
   \end{tabular}
  \end{center}
\caption{SO(9) structure of the B=0 sector of the D=10 SYMQM.
Irreducible representations listed in the last two columns extend over fermionic sectors marked by $\times$ .}
\label{bastab2}
\end{table}

The last example illustrates very well that irreducible representations of SO(9)
stretch across fermionic sectors, which is another way to say that F is not conserved by rotations.
In fact by diagonalizing $J^2$ in other sub-bases, ${\cal B}_{F}(M_{23},M_{45},M_{67},M_{89})$, and matching
eigenvalues among different fermionic sectors, one can construct higher representations and see which $F$'s
they contain. Such a map is shown in Table \ref{bastab2} for few irreps.
The singlet sits only in the middle, $F=12$, sector. Higher irreps
extend gradually towards the edges, i.e. towards the empty and filled fermionic states.
Eventually, beginning with
already discovered (1120), representations span all fermionic sectors. Remember that this structure holds only for
Fock states without bosonic quanta. Since bosonic and fermionic angular momenta add in a usual way,
a singlet with $B=1$, say, would span through $F=10,12$ and $F=14$ sectors, etc. In general, higher B the wider
are irreps in F. Beginning with $B=6$ even simplest irreps stretch across all fermionic sectors.

    Notice however, that each eigenstate of four Cartan generators has well defined $F$ if $B=0$.
In another words, eigenstates of Eqs.(\ref{car}) never stretch across different fermionc sectors,
but irreducible representations do, even for $B=0$. For $B > 0$ the eigenstates of angular momentum
are in general linear combinations of states with different fermion number.

\section{Summary and outlook}

Number of Majorana-Weyl fermions is not conserved by rotations in 9 space dimensions.
This fact has many unusual consequences for the D=10 supersymmetric Yang-Mills quantum mechanics.
Irreducible representations of SO(9) cover many fermionic sectors of the theory.
In particular, the empty Fock state is not a singlet. It belongs to the complicated 132132-dimensional
representation which extends over all sectors with even fermion number. The simplest invariant state
is unique and it is in the half-filled sector with 12 fermions. It is empty with respect to bosonic quanta,
but has quite nontrivial structure in terms of the elementary, gauge invariant,
fermionic Fock states.
It is a linear combination of the 1360 basis states which, out of a total of 56056 states in this sector,
are annihilated by the suitably chosen Cartan generators of SO(9). We have explicitly constructed
this expansion. This state replaces the empty state while building the SO(9) invariant subspaces
of the theory. Therefore one can proceed now with the diagonalization of the hamiltonian in channels with
fixed SO(9) angular momentum.

     We conclude with some open questions, which might help to simplify present solution.
Equation (\ref{car}) for the $J^{89}$ generator suggest that the normal ordering
of Majorana-Weyl creation/annihlation operators might help. However one has to check if this
is consistent with the whole SO(9) algebra and other symmetries of the model.
Second, since the singlet is in the half-filled sector, one wonders if some variant of the Dirac
procedure might work. The trouble is that there are many half-filled sates here and none of the
simple redefinition of $f$'s seems to work. Finally, one might look for the Bogoliubov transformation
which makes our singlet simple. One should keep in mind however, that such a transformation, if exists,
should also render simplicity of the hamiltonian.

\section*{Acknowledgments}
I would like to thank W. A. Bardeen for the discussion.
This work is supported by the Polish Committee for Scientific Research
under the grant no. 1~ P03B~024~27
(2004-2007).

\end{document}